\documentclass[12pt,a4paper,final]{iopart}

\usepackage{iopams}
\usepackage{graphicx}
\usepackage[breaklinks=true,colorlinks=true,linkcolor=blue,urlcolor=blue,citecolor=blue]{hyperref}

\expandafter\let\csname equation*\endcsname\relax
\expandafter\let\csname endequation*\endcsname\relax

\usepackage{float}
\usepackage{color}
\usepackage[usenames,dvipsnames]{xcolor}
\usepackage{mathtools}
\renewcommand{\vec}[1]{\mathbf{#1}}

\usepackage{mathrsfs}

\def\v{p}
\def\Np{N_P}

\def\xd{\dot{x}}
\def\qm{\frac{q}{m}}
\def\xdd{\ddot{x}}
\def\xddd{\dddot{x}}

\def\V{{\cal V}}

\def\Lie{{\cal L}}
\def\G{{\cal G}}
\def\W{{\cal W}}
\def\A{{\cal A}}
\def\X{{\cal X}}
\def\F{{\mathscr F}}
\def\a{\underline{a}}
\def\asquared{\a^2}
\def\uphi{u_\phi}
\def\uxi{u_\xi}
\def\usigma{u_\sigma}

\def\np{401}
\def\SI{19.9}

\def\cp{1197.7}
\def\cs{12.5}
\def\cS{-0.46}
\def\qap{1301.1}
\def\qas{14.4} 
\def\qaS{-0.280} 
\def\qbp{1451.8}
\def\qbs{16.6}
\def\qbS{-0.129}
\def\qcp{1581.3}
\def\qcs{18.0}
\def\qcS{-0.0597}

\begin{document}

\title{Longitudinal and transverse cooling of relativistic electron beams in intense laser pulses}

\author{Samuel~R. Yoffe$^1$,
        Yevgen Kravets$^{2,1}$,
        Adam Noble$^1$
    and Dino~A. Jaroszynski$^1$}

\address{$^1$Department of Physics, SUPA, University of Strathclyde, Glasgow
G4 0NG, UK}
\address{$^2$Centre de Physique Th\'{e}orique,
\'{E}cole Polytechnique, 91120, Palaiseau, France}

\eads{\mailto{d.a.jaroszynski@strath.ac.uk} and \mailto{adam.noble@strath.ac.uk}}

\begin{abstract}
With the emergence in the next few years of a new breed of high power laser
facilities, it is becoming increasingly important to understand how
interacting with intense laser pulses affects the bulk properties of a
relativistic electron beam. A detailed analysis of the radiative cooling of
electrons indicates that, classically, equal contributions to the phase
space contraction occur in the transverse and longitudinal directions. In
the weakly quantum regime, in addition to an overall reduction in beam
cooling, this symmetry is broken, leading to significantly less cooling in
the longitudinal than the transverse directions. By introducing an efficient
new technique for studying the evolution of a particle distribution, we
demonstrate the quantum reduction in beam cooling, and find that it depends
on the distribution of energy in the laser pulse, rather than just the total
energy as in the classical case.
\end{abstract}


\maketitle

\section{Introduction}
The emergence over the next few years of a new generation of ultra-high
power laser facilities, spearheaded by the Extreme Light Infrastructure
(ELI) \cite{url_ELI}, represents a major advance in the possibilities
afforded by laser technology. In addition to important practical
applications, these facilities will, for the first time, allow investigation
of qualitatively new physical regimes. Among the first effects to be
explored will be radiation reaction.

Radiation reaction---the recoil force on an electron due to its emission of
radiation---remains a contentious area of physics after more than a century
of investigation. The standard equation describing radiation reaction (the
so-called LAD equation, after its progenitors Lorentz, Abraham, and Dirac
\cite{Lorentz1916,Abraham1932,Dirac1938})
for a particle of mass $m$ and charge $q$ in an electromagnetic field $F$
reads
\begin{equation}
\label{LAD}
\xdd^a= \frac{f^a_\text{ext}}{m} + \tau \Delta^a{}_b \xddd^b = -\qm F^a{}_b
\xd^b + \tau \big( \xddd^a - \xdd_b \xdd^b \xd^a \big),
\end{equation}
where $f^a_\text{ext} = -q F^a{}_b \xd^b$ is the Lorentz force.
Here, the constant $\tau \coloneqq q^2/6\pi m$ is the `characteristic
time' of the particle\footnote{The characteristic
time $\tau = 2r/3c$ can be interpreted as the time taken for
light to travel across the classical radius of the particle, $r = q^2/4\pi
\epsilon_0 m c^2$. For an electron, $\tau_e = 6.3 \times 10^{-24}$ s,
corresponding to $r_e = 2.8 \times 10^{-15}$ m. Since radiation
damping is proportional to $\tau$, radiation reaction effects will typically
be more prominent for electrons, for which $q=-e$ and $m=m_e$, than for
particles with larger mass.}
and an overdot denotes differentiation with respect to proper
time. Indices are raised and lowered with the metric tensor
$\eta=\text{diag}(-1,1,1,1)$, and repeated indices are summed from 0 to 3.
The $\xd$-orthogonal projection $\Delta^a{}_b \coloneqq
\delta^a_b+\xd^a\xd_b$ ensures that $\xdd$ is orthogonal to $\xd$, preserving the normalisation condition $\xd^a\xd_a=-1$ (equivalently the
mass shell condition, $p^a p_a = -m^2$, where $p^a = m \xd^a = (\gamma
m,\vec{p})$). We work in Heaviside-Lorentz units with $c=1$.

Equation \eqref{LAD} may be unpacked and expressed in terms of the 3-momentum as
\begin{align}
 \label{LL_3d}
 \frac{d\vec{p}}{dt} &= q \left( \vec{E} + \frac{\vec{p}}{\gamma m} \times
 \vec{B} \right) + \tau\gamma \left[ \frac{1}{\gamma} \frac{d}{dt} \left(
 \gamma \frac{d\vec{p}}{dt}
 \right) + \vec{p} \left( \frac{d\gamma}{dt} \right)^2 - \frac{\vec{p}}{m^2} \left( \frac{d\vec{p}}{dt}
 \right)^2 \right] ,
\end{align}
where $\gamma = \sqrt{1 + \vec{p}^2/m^2}$ and $d\gamma/dt = (\vec{p} \cdot d\vec{p}/dt)/\gamma m^2$.

Despite numerous independent derivations of equation \eqref{LAD},
either on the basis of energy-momentum conservation
\cite{Dirac1938,Bhabha1939} or as the Lorentz force due to the particle's
(regularized) self-field \cite{Barut1974}, it is subject to numerous
difficulties; see the recent review \cite{Burton2014} for an account of
these problems and proposed solutions. The most widely used alternative to
LAD is that introduced by Landau and Lifshitz \cite{Landau1962}, by treating
the self-force as a small perturbation about the applied force and retaining
terms to leading order in the small parameter $\tau$:
\begin{align}
 \label{LL}
 \ddot{x}^a = -\frac{q}{m}F^{ab}\dot{x}_b - \tau \frac{q}{m} \left(
 \partial_c F^{ab}\dot{x}_b \dot{x}^c - \frac{q}{m}\Delta^a{}_b F^{bc} F_{cd} \dot{x}^d
\right).
\end{align}
It is often claimed that \eqref{LL} is valid provided only that quantum
effects can be ignored, and though a rigorous demonstration remains elusive
there is mounting evidence that this is indeed the case
\cite{Kravets2013a,Spohn2000}.
Note that equation \eqref{LL} can also be presented in terms of the electric
and magnetic fields, $\vec{E}$ and $\vec{B}$, as
\cite{Landau1962,Tamburini2010}
\begin{align}
 \label{eq:3-LL}
 \frac{d\vec{p}}{dt} &= q \left( \vec{E} + \frac{\vec{p}}{\gamma m} \times
 \vec{B} \right) \nonumber\\
 &\qquad+ \tau\gamma q \left\{ \frac{d\vec{E}}{dt} +
 \left( \frac{\vec{p}}{\gamma m} \times \frac{d\vec{B}}{dt} \right) + \frac{q}{\gamma m}
 \left[ \vec{E} \times \vec{B} + \left( \frac{\vec{p}}{\gamma m} \times
 \vec{B} \right) \times \vec{B} \right]
 \vphantom{\left(\frac{\vec{p}}{\gamma m} \times\vec{B} \right)^2 \vec{p}}
 \right. \nonumber \\
 &\qquad\qquad\qquad \left. + \frac{q}{\gamma^2 m^2}
 \big(\vec{p} \cdot \vec{E} \big) \vec{E} + \frac{q}{\gamma^2
 m^4} \big( \vec{p} \cdot \vec{E} \big)^2 \vec{p} - \frac{q}{m^2} \left( \vec{E} + \frac{\vec{p}}{\gamma m} \times
 \vec{B} \right)^2 \vec{p} \right\},
\end{align}
where the total time derivative acting on the $\vec{E},\vec{B}$ fields is $d/dt = \partial/\partial t +
(1/\gamma m) \vec{p} \cdot \vec{\nabla}$.

Under the conditions expected at ELI, the caveat `provided that quantum
effects can be ignored' is pertinent. Quantum effects are typically
negligible if the electric field observed by the particle, $\hat{E}$, is much less than
the Sauter-Schwinger field \cite{Sauter1931,Schwinger1951} typical of QED
processes, that is provided
\begin{equation}
\chi \coloneqq \frac{e\hbar}{m^2_e}\sqrt{F^{ab}F_{ac}\xd_b\xd^c} =
\frac{\hat{E}}{E_S} \ll 1,
\end{equation}
where $E_S = m_e^2 c^3/e\hbar = 1.32 \times 10^{18}$ V/m is the
Sauter-Schwinger critical field.
For 1 GeV electrons in a laser pulse of intensity $10^{22}$~W/cm$^2$
(parameters obtainable at ELI), $\chi\sim 0.8$ and quantum effects cannot be
ignored. A complete QED treatment of radiation reaction is difficult to
implement and problematic even to define but, provided $\chi$ remains small,
a semi-classical modification to \eqref{LL} should be valid
\cite{Erber1966}.

An important difference between the classical and quantum pictures of
radiation emission can be seen in the radiation spectrum. Classically, a
charged particle can radiate arbitrarily small amounts of energy at all frequencies. However, in the quantum
picture, the particle must radiate entire quanta of energy in the form of
photons. Thus, the energy (frequency) of the emitted photons is limited by the energy of the
particle. This suppresses emission at high frequencies, and introduces a
cut-off in the spectral range of the emitted radiation
\cite{Erber1966}. As such, it is expected that the effects of radiation
reaction are overestimated by classical theories in regimes where quantum
effects become important \cite{Ritus1979}, since they consider the particle
to be radiating at all frequencies.

In order to account for this reduction in the effects of radiation reaction
relative to the Landau--Lifshitz equation of motion, we follow Kirk, Bell and Arka
\cite{Kirk2009} and scale the radiation reaction force by the function
$g(\chi)$:
\begin{align}
 \label{eq:gLL}
 \ddot{x}^a = -\frac{q}{m}F^{ab}\dot{x}_b - g(\chi) \tau \frac{q}{m} \left(
 \partial_c F^{ab}\dot{x}_b \dot{x}^c - \frac{q}{m}\Delta^a{}_b F^{bc} F_{cd} \dot{x}^d
\right).
\end{align}
The full expression for $g(\chi)$ involves a non-trivial integral over
Bessel functions. To make this tractable, we use an approximation introduced
by Thomas \textit{et al.} \cite{Thomas2012},
\begin{equation}
\label{eq:QM_g}
g\left(\chi\right) = \left(1 + 12\chi + 31\chi^2 + 3.7\chi^3\right)^{-4/9}.
\end{equation}
It can be clearly seen that, in the classical limit $\chi \to 0$, we have $g\left(\chi\right)
\to 1$, recovering the classical equation of motion \eqref{LL}. As we move
into a more strongly quantum regime, the quantum nonlinearity parameter
$\chi$ increases and the scaling function $g(\chi)$ decreases, in turn reducing
the effects of radiation reaction.
The model essentially reduces to a rescaling of the characteristic time of
the particle, $\tau \to g(\chi)\tau$, which can also be applied to equation
\eqref{eq:3-LL}. For $\chi \sim 1$, the stochasticity
of quantum emission becomes important, and the semi-classical model is no
longer applicable \cite{Blackburn2014}. At this point, $g(\chi) \simeq
0.18$, which corresponds to a significant reduction in the effects of
radiation reaction.

It is generally accepted that radiation reaction effects will be more
readily observed in the behaviour of particles than in the radiation they
emit \cite{Thomas2012,Ilderton2013}. As such, it is important to be able to
accurately determine the distribution of a bunch of particles evolving
according to \eqref{LL} or its semi-classical extension (\ref{eq:gLL}). Usually this would
involve solving a Vlasov equation \cite{Noble2013a} or following the
evolution of very large numbers of particles \cite{Vranic2014}, either of which is
computationally very intensive.

In this paper we investigate beam cooling of a particle bunch due to
classical and semi-classical models of radiation reaction. In Section
\ref{sec:vlasov} we present a detailed discussion of longitudinal and
transverse phase space contraction of the particle distribution, along with
an analytical solution of the classical Vlasov equation. The longitudinal particle distribution is introduced. Since the semi-classical
Vlasov equation has no analytical solution, in Section \ref{sec:model} we
introduce a new method of accurately reconstructing the particle
distribution from the trajectories of a relatively small number of
particles. Classical predictions using this method are compared to the
analytical solution with excellent agreement. The method is then applied in
Section \ref{sec:results} in order to compare classical and semi-classical
predictions for an electron beam colliding with an intense laser pulse.
Finally, we conclude by summarising our findings in Section
\ref{sec:conclusions}.

\section{Particle distribution and phase space contraction}
\label{sec:vlasov}

The evolution of a particle beam can be described by the Vlasov equation for
the particle distribution $\F(x,u)$, where $u^a =
(\gamma,\vec{u})$ is the 4-velocity. Position and velocity
are considered as independent phase space variables. The Vlasov equation for
$\F$ can be expressed as
\begin{equation}
\frac{d}{ds} \big( \F  \mathbb{V} \big) = \left[ \frac{d\F}{ds} + \beta_s
\F\right] \mathbb{V} =
0,
\end{equation}
where $\mathbb{V}$ is the phase-space volume element
and $\beta_s(x,u)$ describes the rate of change (\textit{i.e.} expansion or contraction) of
$\mathbb{V}$ with proper time $s$. (Technically, $\beta_s$ is the phase-space divergence of the
vector field $X = u^a \partial/\partial x^a + \A^I \partial/\partial
u^I$ (where $\A$ is the acceleration) associated with the flow $d/ds$, given by the Lie derivative
$\Lie_X\mathbb{V} = \beta_s \mathbb{V}$, see \cite{Noble2013a}.) Capital
Latin indices take three values.
Unlike the Liouville equation (or the case with no radiation
reaction) the phase-space volume element is not preserved
by the flow, $\beta_s \neq 0$.

To facilitate investigation of the interaction of a particle bunch with a
laser pulse, we introduce the (null) wavevector $k$ such that the phase of
the pulse is
\begin{equation}
 \label{eq:phase}
 \phi = -k \cdot x = \omega t - \vec{k}\cdot\vec{x}.
\end{equation}
The orthogonal (transverse) vectors $\epsilon,\lambda$ satisfying
\begin{equation}
 \label{eq:basis}
 \epsilon^2 = \lambda^2 = 1 \qquad\text{and}\qquad k \cdot \epsilon 
 = k \cdot \lambda = \epsilon \cdot \lambda = 0,
\end{equation}
together with $k$ and the null vector $\ell$ (defined to satisfy
$\ell\cdot\epsilon = \ell\cdot\lambda = 0$ and $k \cdot \ell = -1$) form a
basis. In addition, the coordinates
\begin{equation}
 \label{eq:coords}
 \xi = \epsilon \cdot x, \qquad \sigma = \lambda \cdot x \qquad \text{and}
 \qquad \psi = -\ell \cdot x
\end{equation}
are also defined, along with the corresponding velocities $\uphi$, $\uxi$,
$\usigma$ and $u_\psi$. However, $u_\psi$ is not independent and may be found from the
normalisation condition $u^a u_a = \uxi^2 + \usigma^2 - 2\uphi u_\psi =
-1$. We note that Greek subscripts are used only
as labels and are not free indices.

For a plane wave with arbitrary polarisation, the electromagnetic field
tensor $F$ depends on spacetime only through the phase $\phi$, and takes the form
\begin{equation}
\label{eq:F_planewave}
\frac{q}{m}F^a{}_b = a_\epsilon(\phi) \big(\epsilon^a k_b - k^a \epsilon_b
\big) + a_\lambda(\phi) \big(\lambda^a k_b - k^a \lambda_b \big),
\end{equation}
where the functions $a_{\epsilon,\lambda}(\phi)$ are dimensionless measures
of the electric field strength in the $\epsilon$,$\lambda$ direction. The
corresponding electric and magnetic fields are
$\vec{E} = (m \omega/q) [ a_\epsilon(\phi) \boldsymbol{\hat{\epsilon}} +
a_\lambda(\phi) \boldsymbol{\hat{\lambda}} ]$ and $\vec{B} = \vec{k} \times
\vec{E}/\omega$, where the orthogonal unit 3-vectors
$\boldsymbol{\hat{\epsilon}},\boldsymbol{\hat{\lambda}}$ satisfy $\vec{k} \cdot \boldsymbol{\hat{\epsilon}} = \vec{k}
\cdot \boldsymbol{\hat{\lambda}} = 0$.

In a similar manner, we assume that the particle distribution also depends
on spacetime only through the phase $\phi$, such that $\F(x,u) =
\F(\phi,\uphi,\uxi,\usigma)$. The Vlasov equation is then written
\begin{equation}
\label{eq:Vlasov:general}
\underbrace{\uphi\frac{\partial \F}{\partial \phi}+ \A^I\frac{\partial \F}{\partial
u^I}}_{d\F/ds}+ \underbrace{\uphi\frac{\partial}{\partial
u^I}\left(\frac{\A^I}{\uphi}\right)}_{\beta_s} \F = 0,
\end{equation}
where $u^I\in \{\uphi,\uxi,\usigma\}$, and the accelerations $\A^I \in
\{\A_\phi,\A_\xi,\A_\sigma\}$ follow from the single-particle equations of
motion. Dividing through by $\uphi$, we have
\begin{equation}
 \label{eq:Vlasov_phi}
 \frac{d\F}{d\phi} + \beta\F = 0, \qquad \text{where} \qquad \beta =
 \frac{\partial}{\partial u^I} \left( \frac{\A^I}{\uphi} \right).
\end{equation}
The quantity $\beta$ is responsible for any phase space contraction ($\beta
< 0$) or expansion ($\beta > 0$) of the particle distribution, and the associated change in electron entropy \cite{Burton2014b}.

For a highly relativistic particle beam colliding with a laser pulse (the
scenario in which radiation reaction effects are most prominent), we are mainly interested
in the dependence of $\F$ on $\phi$ and $\uphi$. An advantage of the
coordinate system \eqref{eq:phase}--\eqref{eq:coords} is that it decouples the longitudinal
from the transverse velocity in the Lorentz invariant measure, $d^3\xd/\gamma = d\uxi d\usigma d\uphi/\uphi$. Hence we can
define the \emph{longitudinal distribution}
\begin{equation}
 f(\phi,\uphi) = \int_{\mathbb{R}^2} \F\ d\uxi d\usigma,
\end{equation}
which satisfies the \emph{reduced Vlasov equation}
\begin{equation}
 \label{eq:Vlasov_reduced}
 \frac{df}{d\phi} + \beta_\parallel f = 0, \qquad \text{where} \qquad
 \beta_\parallel = \frac{\partial}{\partial \uphi} \left(
 \frac{\A_\phi}{\uphi} \right).
\end{equation}
Here, $\beta_\parallel$ describes the \emph{longitudinal} phase space contraction. The transverse
contribution is then
\begin{equation}
 \label{eq:beta_perp}
 \beta_\perp = \beta - \beta_\parallel = \frac{\partial}{\partial \uxi} \left(
 \frac{\A_\xi}{\uphi} \right) + \frac{\partial}{\partial \usigma} \left(
 \frac{\A_\sigma}{\uphi} \right).
\end{equation}
Note that this reduction to the longitudinal distribution is purely a
consequence of the coordinate system, and does not rely on the plane wave
assumption.

It is at this point that a decision must be made as to the appropriate
single-particle equations of motion. While there are many classical models
for radiation reaction \cite{Burton2014}, we start by considering the
Landau--Lifshitz equation given by \eqref{LL}, before moving on to the
semi-classical extension \eqref{eq:gLL}. This is in part motivated by
the existence of an analytical solution to the single-particle
Landau--Lifshitz equation \cite{DiPiazza2008}.
In our coordinates, the Landau--Lifshitz equations in the plane wave
\eqref{eq:F_planewave} are
\begin{align}
\label{eq:LL_coords}
\hat{\A}_\phi &= -\tau \asquared \uphi^3 \nonumber \\
\hat{\A}_\xi  &= - \uphi \big( a_\epsilon + \tau \uphi a'_\epsilon \big) - \tau
\asquared \uphi^2 \uxi \\
\hat{\A}_\sigma &= - \uphi \big( a_\lambda + \tau \uphi a'_\lambda \big) - \tau
\asquared \uphi^2 \usigma, \nonumber
\end{align}
where $\asquared(\phi) = a^2_\epsilon(\phi) + a^2_\lambda(\phi)$ and prime denotes differentiation
with respect to $\phi$. Inserting these equations into
\eqref{eq:Vlasov_reduced} and \eqref{eq:beta_perp}, we find for the
classical case
\begin{equation}
 \label{eq:beta_classical}
 \hat{\beta}_\parallel = \hat{\beta}_\perp = -2\tau \asquared \uphi \leq 0.
\end{equation}
It is immediately apparent that half the contraction of the distribution
occurs in the longitudinal and half in the transverse directions.

\begin{figure}[tb]
 \centering
 \includegraphics[width=0.49\textwidth]{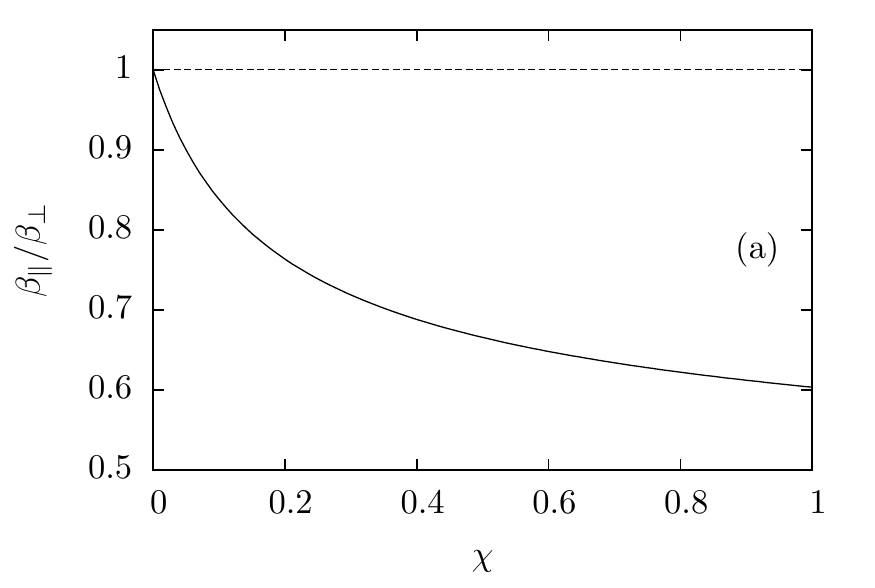} \hfill
 \includegraphics[width=0.49\textwidth]{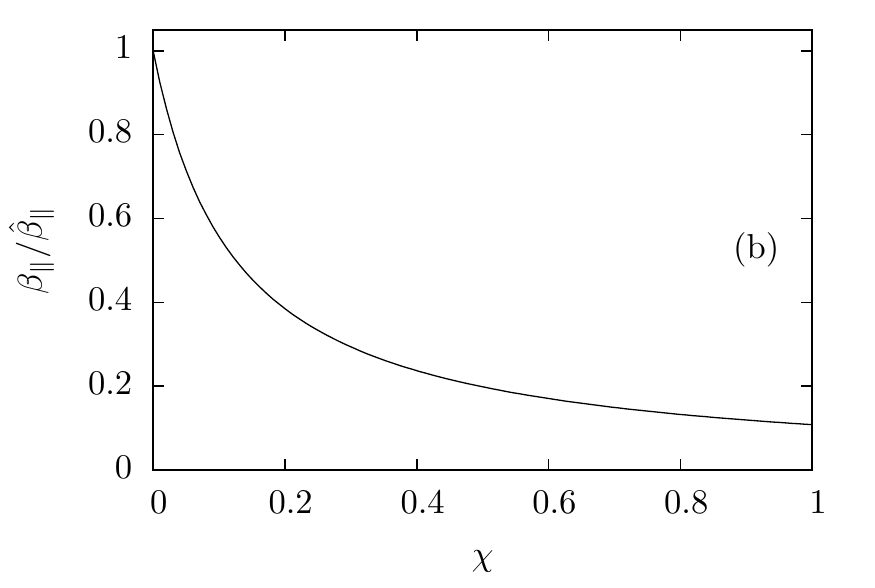}
 \caption{Reduction of longitudinal beam cooling. Part (a): the ratio
 \eqref{eq:contr_ratio} of the longitudinal to the transverse phase space
 contraction in the semi-classical model. The dashed line shows the
 classical ratio $\hat{\beta}_\parallel/\hat{\beta}_\perp = 1$. Part (b): ratio of the semi-classical
 longitudinal contraction to the classical Landau--Lifshitz result.}
 \label{fig:contraction}
\end{figure}

The semi-classical equations of motion are just
\eqref{eq:LL_coords} with the replacement $\tau \to g(\chi)\tau$. However, since
$\chi(\phi,\uphi) =
3\tau\a(\phi)\uphi/2\alpha$ (where $\alpha$ is the fine structure constant) depends on $\uphi$ (but not on the transverse velocities) we pick up an
additional contribution to the longitudinal phase space contraction:
\begin{align}
 \beta = g
 \hat{\beta} + \frac{\partial g}{\partial \uphi} \frac{\hat{\A}_\phi}{\uphi} \qquad
 \text{and} \qquad \beta_\parallel = g \hat{\beta}_\parallel +
 \frac{\partial g}{\partial \uphi} \frac{\hat{\A}_\phi}{\uphi};
\end{align}
whereas, the transverse contraction is simply scaled by $g(\chi)$:
\begin{equation}
 \beta_\perp = \beta - \beta_\parallel = g(\hat{\beta} -
 \hat{\beta}_\parallel) = g \hat{\beta}_\perp = g\hat{\beta}_\parallel.
\end{equation}
Thus, as quantum effects become more important and $g(\chi)$ decreases, the
semi-classical model predicts a reduction in both the longitudinal and
transverse phase space contraction (reduced beam cooling). As well as this
scaling of the classical contraction by $g(\chi)$, there is an additional
\emph{longitudinal heating} given by
\begin{align}
 \frac{\partial g}{\partial\uphi} \frac{\hat{\A}_\phi}{\uphi} =
 \frac{dg}{d\chi} \frac{\partial \chi}{\partial\uphi}
 \frac{\hat{\A}_\phi}{\uphi}
 &= -\beta_\perp \left[ \frac{2}{9} \chi g^{9/4}(\chi) \big( 12 + 62
 \chi + 11.1 \chi^2 \big) \right] \geq 0,
\end{align}
where $\beta_\perp = g\hat{\beta}_\perp = -2\tau\asquared(\phi) g(\chi) \uphi = -4
\alpha\a(\phi)\chi g(\chi)/3$. The ratio
\begin{align}
 \frac{\beta_\parallel}{\beta_\perp} &= 1 + \frac{1}{2} \frac{d
 \log{g}}{d\log{\chi}} \nonumber \\
 \label{eq:contr_ratio}
 &= 1 - \frac{2}{9} \chi g^{9/4}(\chi) \big( 12 + 62
 \chi + 11.1 \chi^2 \big) \leq 1
\end{align}
measures the strength of the longitudinal compared to the transverse phase
space contraction. This is shown in Fig. \ref{fig:contraction}(a) for the interval $\chi \in
[0,1]$. Even for the weakly quantum regime in which the semi-classical model
remains valid, we observe a significant reduction in longitudinal beam
cooling. This is especially clear when comparison is made with the classical
result $\hat{\beta}_\parallel$ as shown in Fig. \ref{fig:contraction}(b). We
see that where $\chi = 0.2$ there is nearly a 60\% reduction in the
longitudinal contraction experienced compared to the Landau--Lifshitz model.

For the case of the classical Landau--Lifshitz theory in a plane wave, the
Vlasov equation \eqref{eq:Vlasov_phi} may be solved analytically for the
particle distribution:
\begin{equation}
 \F(\phi,\uphi,\uxi,\usigma) =
 \F\left(\phi^0,\uphi^0,\uxi^0,\usigma^0\right)
 e^{4\Lambda(\phi,\uphi)},
\end{equation}
where $\left\{ \phi^0, \uphi^0,\uxi^0,\usigma^0 \right\}$ are the
\emph{initial} phase and velocities of a particle with $\{
  \uphi,\uxi,\usigma \}$
at phase $\phi$. In a similar manner, the longitudinal distribution is found to
be
\begin{equation}
 f(\phi,\uphi) = f\left(\phi^0,\uphi^0\right) e^{2\Lambda(\phi,\uphi)}.
\end{equation}
The contraction/expansion of phase space is contained in the function
\begin{equation}
 \label{eq:Lambda}
 \Lambda(\phi,\uphi) = \tau \int_{\phi^0}^\phi d\vartheta\ \asquared(\vartheta)
 \uphi(\vartheta).
\end{equation}
Solutions to equation \eqref{eq:LL_coords} \cite{DiPiazza2008} can then be used to rewrite
$\left\{ \phi^0, \uphi^0,\uxi^0,\usigma^0 \right\}$ in terms of the independent variables $\{\phi, \uphi,\uxi,\usigma \}$:
\begin{align}
 \uphi^0 &= \frac{\uphi}{1 - \tau \uphi \G(\phi)} \nonumber \\
 \label{eq:solve_LL}
 \uxi^0  &= \frac{\uxi - \tau \uphi \Big( \W_\epsilon(\phi) +
 a_\epsilon(\phi) - a_\epsilon(\phi^0) \Big)}{1 - \tau \uphi \G(\phi)} -
 \V_\epsilon(\phi) \\
 \usigma^0  &= \frac{\usigma - \tau \uphi \Big( \W_\lambda(\phi) +
 a_\lambda(\phi) - a_\lambda(\phi^0) \Big)}{1 - \tau \uphi \G(\phi)} -
 \V_\lambda(\phi) , \nonumber
\end{align}
where the functions
\begin{align}
 \label{eq:sol_funcs}
 \G(\phi)   &= \int_{\phi^0}^\phi d\vartheta\ \asquared(\vartheta) \nonumber \\
 \V_i(\phi) &= \int_{\phi^0}^\phi d\vartheta\ a_i(\vartheta) \\
 \W_i(\phi) &= \int_{\phi^0}^\phi d\vartheta\ a_i(\vartheta) \G(\vartheta) \nonumber
\end{align}
with $i \in \{ \epsilon, \lambda \}$ depend only on the properties of the
laser pulse.

Using equations \eqref{eq:solve_LL} we can express $\uphi(\vartheta)$ in
equation \eqref{eq:Lambda} in terms of the independent variable $\uphi$,
\begin{align}
 \Lambda(\phi,\uphi) &= \tau \uphi \int_{\phi^0}^\phi d\vartheta\ 
 \frac{\asquared(\vartheta)}{1 - \tau \uphi \big(\G(\phi) - \G(\vartheta) \big)}
 \nonumber \\
 &= - \ln \Big( 1 - \tau \uphi \G(\phi) \Big),
\end{align}
such that the distribution becomes
\begin{equation}
 \label{eq:solution}
 \F(\phi,\uphi,\uxi,\usigma) =
 \frac{\F\left(\phi^0,\uphi^0,\uxi^0,\usigma^0\right)}{\big(1 - \tau
 \uphi \G(\phi)\big)^4},
\end{equation}
or for the longitudinal distribution
\begin{equation}
 \label{eq:reduced_solution}
 f(\phi,\uphi)
 = \frac{f\left(\phi^0,\uphi^0\right)}{\big(1 - \tau \uphi \G(\phi)\big)^2}
 = \frac{f\left(\phi^0,\frac{\uphi}{{1 - \tau \uphi
 \G(\phi)}}\right)}{\big(1 - \tau \uphi
 \G(\phi)\big)^2}.
\end{equation}
This latter result is in agreement with observations made by Neitz and Di~Piazza
\cite{Neitz2014}, and we see that the longitudinal distribution is only sensitive to the
properties of the laser pulse through the function $\G(\phi)$. After the
pulse has passed, $\G$ becomes constant and
is proportional to the fluence of the pulse. Final-state
properties of the longitudinal distribution therefore depend only on the total
energy contained in the laser pulse, and are insensitive to how that energy
is distributed within the pulse. The full distribution $\F$, on the
other hand, depends additionally on the integrals $\W_i$ given in equation
\eqref{eq:sol_funcs}.

Although the reduced Vlasov solution \eqref{eq:reduced_solution} is somewhat simpler
than \eqref{eq:solution}, and captures the key features of the electron beam
itself, the solution is not sufficient to calculate the transverse current
density, and hence cannot be coupled to Maxwell's equations to determine the
radiation produced by the electron beam. However, if the transverse momentum
spread is sufficiently small, we can approximate the full distribution by
\begin{equation}
 \F(\phi,\uphi,\uxi,\usigma) = f(\phi,\uphi)
 \delta\left(\uxi - \X_\epsilon(\phi,\uphi) \right)
 \delta\left(\usigma - \X_\lambda(\phi,\uphi) \right),
\end{equation}
where the $\delta$-functions restrict the transverse velocities to the
submanifold $\X_i$.
Then, in addition to \eqref{eq:Vlasov_reduced}, equation
\eqref{eq:Vlasov_phi} yields
\begin{align}
 \label{eq:reduced_X}
 \frac{\partial \X_i}{\partial \phi} - \tau\asquared\uphi^2
  \frac{\partial \X_i}{\partial \uphi} & = -\left(a_i +
  \tau\uphi a^\prime_i + \tau\uphi\asquared\X_i \right), \qquad \text{for}\
  i \in \{\epsilon,\lambda\}.
\end{align}
Note that \eqref{eq:reduced_X} indicate that the distribution is
concentrated on a surface in phase space that itself satisfies the
Landau--Lifshitz equation.

Given solutions to the reduced Vlasov equation \eqref{eq:reduced_solution} and the
transverse Landau--Lifshitz equation \eqref{eq:reduced_X}, the current can
be written
\begin{equation}
 j^a = q\int \F \xd^a\ d\uxi d\usigma \frac{d\uphi}{\uphi}
     = j^a_\perp + \varrho\ell^a + j^a_\parallel,
\end{equation}
with $j^a_\perp$ and $\varrho$ evaluated as
\begin{equation}
 j^a_\perp = q\epsilon^a\int f\ \X_\epsilon \frac{d\uphi}{\uphi} +
 q\lambda^a\int f\ \X_\lambda \frac{d\uphi}{\uphi} \qquad \text{and} \qquad
 \varrho = q\int fd\uphi\ .
\end{equation}
We could also calculate $j^a_\parallel$ directly, but it follows more
straightforwardly from charge conservation, $\partial_a j^a = 0$.

In the following, we restrict our attention to the longitudinal distribution
$f(\phi,\uphi)$, and longitudinal beam cooling, as this is more readily
measurable in experiments than the transverse cooling. However, the
transverse cooling, which can be considerably greater, can be determined
from equation \eqref{eq:contr_ratio}.

\section{Numerical (re)construction of the particle distribution}
\label{sec:model}

The motion of a single charged particle colliding with a laser pulse,
including radiation reaction, has been extensively studied
\cite{Kravets2013a,DiPiazza2008,Lehmann2011,Harvey2011}. As shown in Section
\ref{sec:vlasov}, the Vlasov equation with classical readiation reaction can be solved analytically. However, this is not the case for the
semi-classical model \eqref{eq:gLL} or for stochastic models of
radiation reaction in the quantum regime. Instead of attempting to solve a Vlasov-type
equation on the phase space numerically, which would require significant
computing resources, we propose an innovative method which allows for the
dynamics of a particle \emph{distribution} to be explored using
single-particle equations of motion such that the distribution can be
efficiently reconstructed. While this approach is quite general and could be
used for a variety of systems, here we consider a distribution of particles
subject to equation \eqref{LL} and its semi-classical extension \eqref{eq:gLL}, without
particle-particle interactions\footnote{For a highly relativistic particle
bunch, these interactions can be neglected on the time scale of the laser
interaction.}.

Assuming that the laser pulse can be approximated by a plane wave with
compact longitudinal support\footnote{A function has compact support if it
is zero outside a finite interval.}, any spatial spread in the initial
particle distribution would only determine the moment when each particular
particle enters the pulse. For simplicity, we therefore take all particles
to originate from the same point. This is reasonable as we are primarily
interested in the longitudinal momentum distribution.

Since our pulse is modelled by a plane wave and we focus on the longitudinal
properties of the distribution, we consider the initial momenta to be
strongly peaked about zero in the transverse directions. As such, the
initial distribution can be taken to be a Maxwellian distribution for the
(longitudinal) momentum $\v$ (in units of $mc$)
\begin{equation}
 \label{eq:initial_dist}
 f\left(\phi=0,\v\right) = \frac{\Np}{\sqrt{2\pi\theta}}
 \exp\left[-\frac{(\v-\bar{\v})^2}{2\theta}\right],
\end{equation}
with $\phi = \omega t - \vec{k}\cdot\vec{x}$ the phase, $\theta$ the
variance of the distribution, and $\Np$ the number of particles. The
momentum $p$ is related to our velocities of Section \ref{sec:vlasov} by
\begin{equation}
 \label{eq:mom_coords}
 p = \uphi/\omega-\gamma, \qquad \text{where} \qquad \gamma = \frac{1 + 
 \uxi^2 + \usigma^2+(\uphi/\omega)^2}{2(\uphi/\omega)}.
\end{equation}
We stress
that this initial distribution is chosen for its simplicity; alternative
distributions could be used where appropriate (such as Maxwell--J\"uttner).

Typically, one would sample the distribution at random, which would require
a large number of particles to accurately represent the distribution.
Instead, since the particle number is simply
\begin{equation}
\label{eq:number_density}
 \Np = \int_{-\infty}^{\infty} d\v\ f\left(\phi,\v\right),
\end{equation}
we determine the momentum spacing $\delta \v$ between the particles from the
initial distribution by truncating the integral in \eqref{eq:number_density}
so that the particle number increases by unity in the given momentum
interval:
\begin{equation}
\label{eq:momentum_interval}
 1 = \int\limits_{\v-\frac{\delta \v}{2}}^{\v+\frac{\delta \v}{2}}
 d\v\ f\left(0,\v\right) \simeq f\left(0,\v\right)\delta \v.
\end{equation}
This leads to a set of $\Np = 2N_c + 1$ initial momenta $V(0) =
\big\{\v_i(0)\big\}$ for $i \in [-N_c,N_c]$, with the $\v_i$ generated
iteratively from $\v_0 = \bar{\v}$ and $\v_{\pm 1} = \bar{\v} \pm
1/f(0,\bar{\v})$ using
\begin{equation}
 \label{eq:initial_p}
 \v_i = \v_{i-2\xi} + \frac{2\xi}{f(0,\v_{i - \xi})}\ \quad\text{with}\quad 
 \xi = \text{sgn}(i).
\end{equation}
The momentum space is not sampled uniformly, instead more particles are
located in regions where the distribution function is large.

As the evolution proceeds, this procedure is applied in reverse to
reconstruct the distribution. The set of momenta $V(\phi)$ is ordered such
that $\v_{i+1} \ge \v_{i}$ and used to find $\delta \v_i(\phi) =
\big(\v_{i+1}(\phi) - \v_{i-1}(\phi)\big)/2$. The velocity distribution is
then defined to be
\begin{equation}
 \label{eq:reconstruct_f}
 f(\phi,\v_i) \coloneqq \frac{1}{\delta \v_i(\phi)}.
\end{equation}
Reconstruction of a distribution from a particle sample can be problematic,
but in our formalism it becomes quite natural. This is achieved by
using the momentum spacing between particles to determine the value of the
distribution such that equation \eqref{eq:momentum_interval} is satisfied for all
$\phi$ (\textit{i.e.} integration over each of the measured momentum
spacings always contributes a single
particle to the total particle number). The closer the measured momenta are
together, the `more likely' one is to have a particle in that
momentum range, resulting in a larger value for the distribution.

The definition \eqref{eq:reconstruct_f} allows properties of the
distribution to be calculated directly from the momenta of the individual
particles. The mean is simply evaluated as
\begin{align}
 \bar{\v}(\phi) &= \langle \v(\phi) \rangle = \frac{1}{\Np} \int d\v\ \v f(\phi,\v) \nonumber \\
 &\simeq \frac{1}{\Np} \sum_i \v_i(\phi) \
 \underbrace{f(\phi,\v_i)\delta\v_i(\phi)}_{=1}
 \nonumber \\
 \label{eq:mean_p}
 &= \frac{1}{\Np} \sum_i p_i(\phi).
\end{align}
In a similar manner, higher-order moments of the distribution $X_n$ may be
calculated straightforwardly as:
\begin{align}
 \label{eq:moment_p}
 X_n(\phi) = \left\langle \big[\v-\bar{\v}(\phi)\big]^n \right\rangle 
     \simeq \frac{1}{\Np} \sum_i \big[p_i(\phi) - \bar{\v}(\phi)\big]^n.
\end{align}
For example, the variance $\theta(\phi) = X_2(\phi)$ and skewness $S(\phi) =
X_3(\phi)/X_2^{3/2}(\phi)$. We note that angular brackets $\langle \cdots
\rangle$ will be used to denote an average over the bunch, not time.

\begin{figure}[tb]
 \centering
 \includegraphics[width=0.49\textwidth]{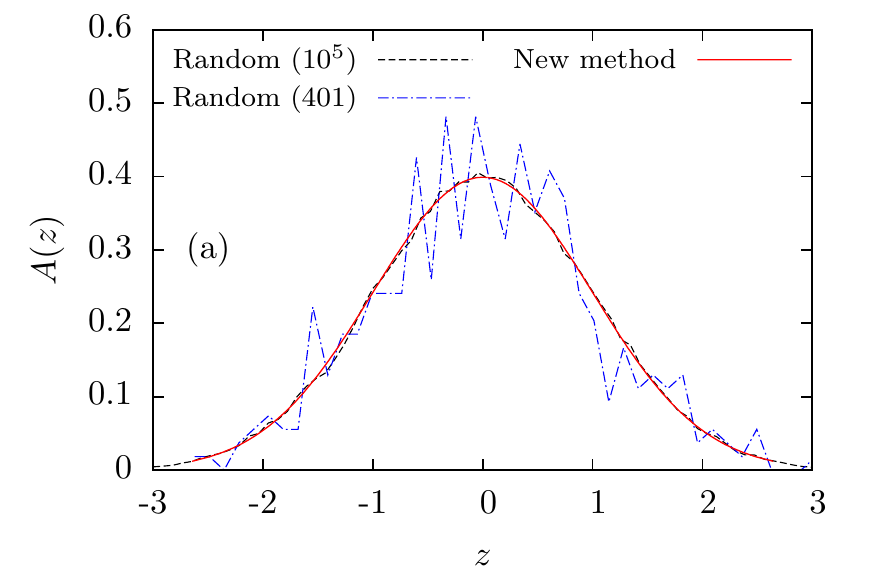} \hfill
 \includegraphics[width=0.49\textwidth]{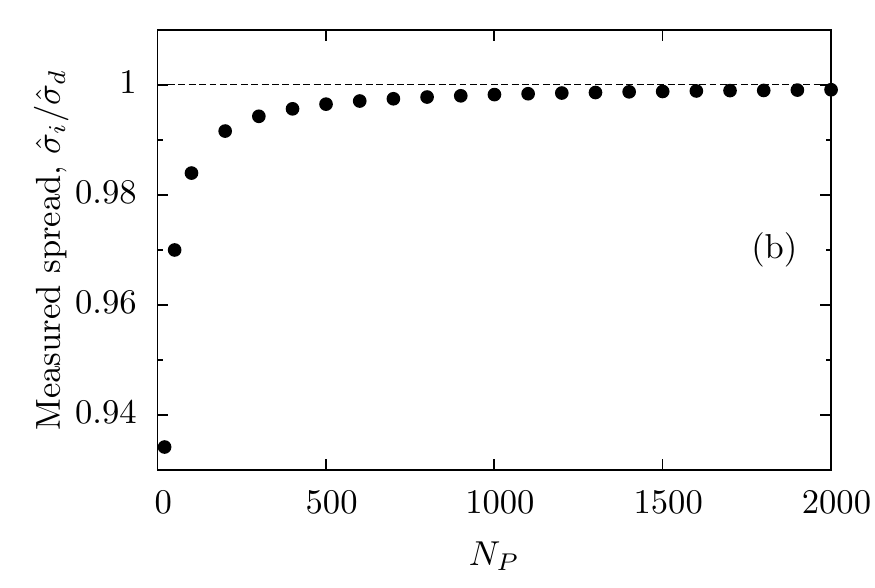}
 \caption{(Colour online) Part~(a): A Gaussian distribution $A(z)$ is
 reconstructed using the new method
 described by equations \eqref{eq:initial_p} and \eqref{eq:reconstruct_f}
 with $\Np = 401$ particles (red, solid) and compared to the standard
 approach of random sampling, using
 both $N_z = 401$ (blue, dot-dash) and $N_z = 100,000$ (black, dashed) particles. Part~(b): Variation
 of the measurement of the initial relative momentum spread $\hat{\sigma}_i$
 as the number of particles $\Np$ is increased, compared to the desired
 value, $\hat{\sigma}_d$.}
 \label{fig:recreate_dist}
\end{figure}

A major advantage of this new methodology is that the distribution can be
accurately represented and reconstructed using far fewer particles than
would be required using random sampling. Figure~\ref{fig:recreate_dist}(a)
shows a simple Gaussian distribution with zero mean and unit
variance, $A(z) = (1/\sqrt{2\pi}) \exp(-z^2/2)$, reconstructed using
two methods. First, the distribution $A(z)$ was sampled according to the
iterative relation \eqref{eq:initial_p} with $\Np=401$ and then reconstructed using
\eqref{eq:reconstruct_f} ({\color{red}\textbf{---}}).
Next, $N_z = 100,000$ random numbers were
generated from $A(z)$ using a Mersenne-Twister pseudo-random number
generator, from which the distribution was found using 100 fixed-width bins
(-~-~-). Even with 100,000 random samples,
the standard approach does not describe the Gaussian perfectly,
while the new method presented in this paper
does an excellent job using only 401 points. For comparison, a sample of
$N_z = 401$ was also reconstructed using 50 fixed-width bins
({\color{blue}--~$\cdot$~--}). The advantage of the new method with such a
small sample size is clear.

Figure~\ref{fig:recreate_dist}(a) also shows how sampling the distribution
according to equation \eqref{eq:initial_p} cuts off the tails of the
distribution, which will have an effect on the measured properties of the
reconstructed distribution. The finite number of particles used to
represent the distribution causes the measured relative momentum spread
(defined by equation \eqref{eq:properties} below) to be less than that
specified when defining the initial distribution. Essentially, it comes down to the `$\simeq$' in equation
\eqref{eq:momentum_interval}, compared to the definition given by equation
\eqref{eq:reconstruct_f}. As $\Np$ is increased, the approximation in
equation \eqref{eq:momentum_interval} improves, and the measured value
approaches the desired value. In addition, with more particles the
distribution is sampled further into the tails. Figure~\ref{fig:recreate_dist}(b) shows the measured
initial spread, $\hat{\sigma}_i$, as a fraction of the desired spread,
$\hat{\sigma}_d$, when the particle number is varied from as few as 11 up to
$\Np=2001$. We see that the approximation quickly improves as $\Np$ is
increased up to about 500. The value $\Np = \np$ is chosen to give less than
0.5\% error in the initial measured momentum spread. In practice, good
agreement can be found with lower $\Np$, with the caveat that properties
sensitive to the tails of the distribution (such as the skewness) may be
strongly affected. (This is confirmed in Fig. \ref{fig:analytical}, where
the distribution and its statistics calculated from the analytical solution to
the classical Vlasov equation are compared to those obtained with this new
method.)

\section{Interaction of a particle bunch with high-fluence laser
pulses}
\label{sec:results}
The analytical solution to the Vlasov equation including radiation reaction
according to the Landau--Lifshitz theory derived in Section \ref{sec:vlasov}
predicts that the collision of an energetic electron beam with a high-intensity
laser pulse leads to a significant contraction of the particle phase space,
resulting in a reduction in the relative momentum spread. Agreement between this analytical
solution and numerical results obtained using the approach discussed above
is shown below to be excellent. As previously observed, classical beam cooling depends only on
the total fluence of the pulse, rather than its duration or peak intensity
independently \cite{Neitz2014,Kravets2014}. However, for the semi-classical
extension, the Vlasov equation is no longer tractable. This highlights the
value of our approach and, to demonstrate the use of our proposed method
in such a case, we consider the importance of quantum effects in the
interaction of an electron bunch with a high-intensity laser pulse.

To establish the impact of quantum effects on the evolution of the particle
distribution subject to radiation reaction, we introduce the \emph{relative
momentum spread} and the \emph{momentum skewness} (calculated from the mean
$\bar{\v}$ and variance $\theta$):
\begin{align}
 \label{eq:properties}
 \hat{\sigma}(\phi) = \frac{\sqrt{\theta(\phi)}}{\bar{\v}(\phi)} \quad
 \text{and} \quad S(\phi) = \frac{\left\langle \big[\v-\bar{\v}(\phi)\big]^3
 \right\rangle}{\theta^{3/2}(\phi)}.
\end{align}
The former gives a measure of the beam quality, while the latter indicates
how symmetric the distribution is about its mean.

We restrict our attention to a linearly polarised $N$-cycle plane wave pulse
\eqref{eq:F_planewave}, modulated by
a $\sin^2$-envelope \cite{Kravets2013a}, with $a_\lambda=0$ and $a_\epsilon=a(\phi)$, where
\begin{equation}
 \label{eq:pulse}
 a(\phi) = \left\{
   \begin{array}{ll}
     a_0 \sin(\phi) \sin^2\left(\pi \phi / L \right) & \text{for }0 < \phi
     < L \\
     0 & \text{otherwise}
   \end{array} \right. ,
\end{equation}
where $a_0$ is the dimensionless (peak) intensity parameter (the so-called
``normalised vector potential'') and $L = 2\pi N$ is the pulse
length\footnote{For the $\sin^2$-envelope used in this work, the full-width
half-maximum (FWHM) duration is half of this value.}. This pulse shape
offers compact support, allowing the particles to begin and end in vacuum.
The total fluence (energy per unit area) of the pulse is proportional to
\begin{equation}
 \mathcal{E} = \int_0^L d\phi\ a^2(\phi) = \frac{3\pi}{8} N a_0^2.
\end{equation}
In this work, $\mathcal{E}$ is kept constant, which fixes $a_0$ for each
$N$. It has been shown \cite{Neitz2014,Kravets2014} (see also Section \ref{sec:vlasov}) that the classical
Landau--Lifshitz prediction for the final state of a particle distribution
emerging from the pulse is completely determined by the fluence, whereas
quantum effects are expected to depend on the value of $a_0$ itself. We are
then able to explore the impact of the reduced emission in the quantum model
with varying $a_0$ while maintaining the same classical prediction. This
allows us to explore the \emph{relative} importance of quantum effects.

To motivate this study, parameters have been chosen to be relevant at the
forthcoming ELI facility. We have chosen to consider $N a_0^2 =
9.248\times10^{3}$ which, for $N=20$ with a wavelength of $\lambda = 800$
nm, represents a full-width half-maximum pulse duration of $27$~fs with
\emph{peak}\footnote{Peak intensity is obtained from $I_\text{peak} =
(4\pi^2 m_e^2 c^5 \epsilon_0 / e^2 \lambda^2) a_0^2 \simeq 2.74 \times
10^{10} (a_0/\lambda)^2$ W/m$^2$.} intensity $2\times10^{21}$~W/cm$^2$. We
have investigated pulses of length $N \in [5,200]$ cycles (together with
their corresponding $a_0$) counter-propagating relative to a bunch of $\Np =
\np$ particles, with an initial momentum spread of $20\%$ around $\sqrt{1 +
\bar{p}^2} = 2\times 10^3$. This corresponds to an average particle energy
of just over 1 GeV, which should be well within the capabilities of the
laser-plasma wakefield accelerator at ELI.

\begin{figure*}[tb]
 \begin{center}
  \includegraphics[width=\textwidth, clip=true, trim=0 70 0 0]
  {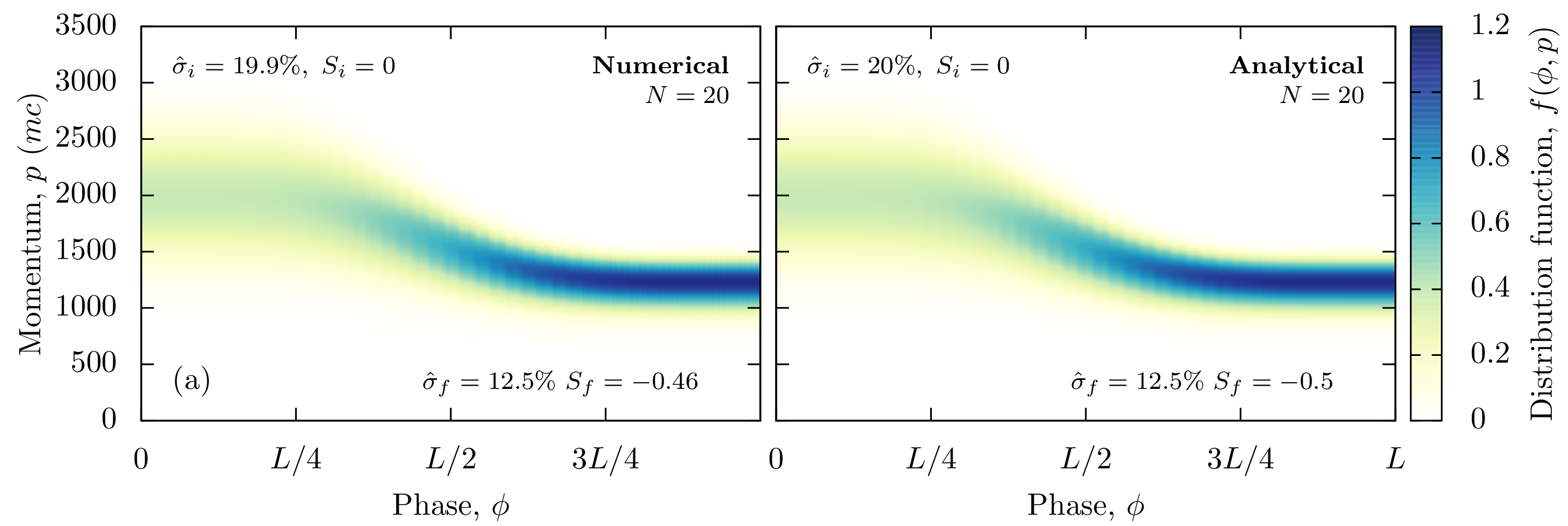}\\
  \vspace*{1em}
  \includegraphics[width=\textwidth]
  {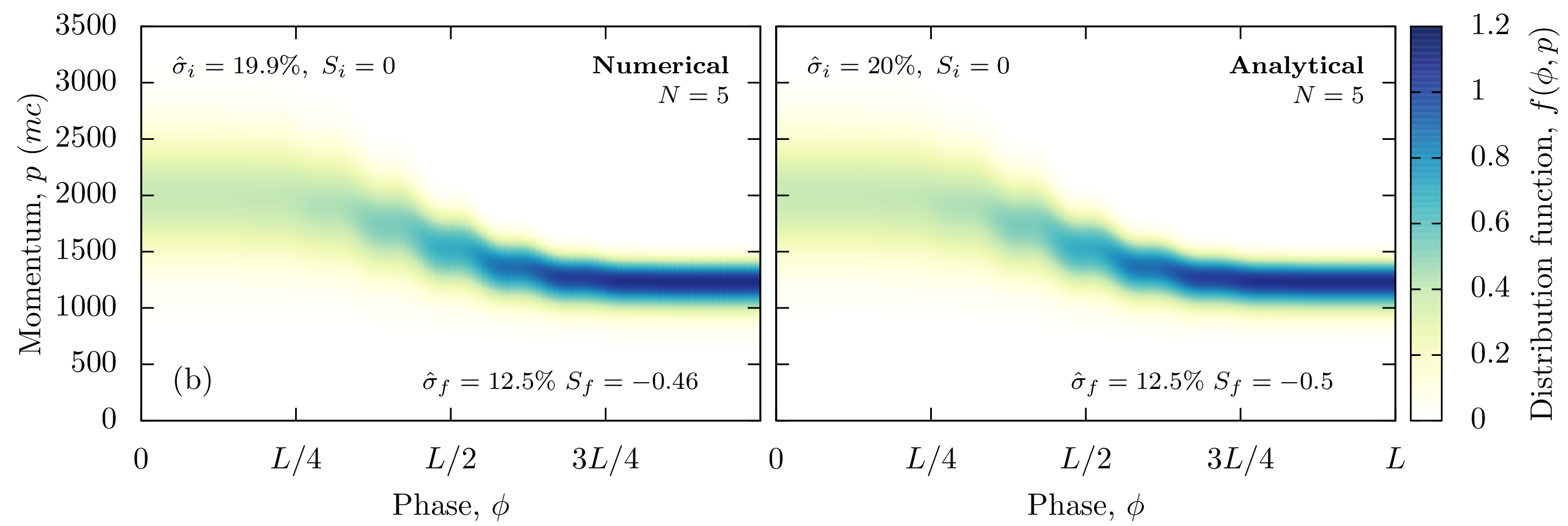}
 \end{center}
 \caption{(Colour online) The phase space evolution of the distribution
 function $f(\phi,p)$ predicted by the analytical solution
 \eqref{eq:reduced_solution} of the reduced Vlasov equation
 \eqref{eq:Vlasov_reduced}, compared to numerical results obtained using the
 new method presented in Section \ref{sec:model}. Results are presented for (a) $N
 = 20$ and (b) $N = 5$ cycles, where $L = 2\pi N$ is the pulse length. The fluence has
 been kept constant, with $N a_0^2 = 9.248 \times 10^3$.}
 \label{fig:analytical}
\end{figure*}

Before comparing predictions of the classical and semi-classical
models, we briefly confirm the validity of our method by comparing numerical
results with the analytical solution
\eqref{eq:reduced_solution} obtained in Section \ref{sec:vlasov}. Figure
\ref{fig:analytical} shows the interaction of a
1 GeV electron beam with a plane-wave laser for two pulse lengths, $N
= 20$ ($a_0 \simeq 22$) in part (a) and $N = 5$ ($a_0 \simeq 43$) in part
(b). The left-hand panels show the numerical results obtained using the new
method described in Section \ref{sec:model}, while the right-hand panels
show the solution \eqref{eq:reduced_solution} using the initial
distribution $f(0,\uphi)$ corresponding to $f(0,\v)$ given by
\eqref{eq:initial_dist} with $\v =\frac{1}{2}(\uphi/\omega-\omega/\uphi)$. The momentum
$p$ is evaluated during the evolution using \eqref{eq:mom_coords}, along
with the solutions \eqref{eq:solve_LL} for $\uxi,\usigma$ when $\uxi^0 =
\usigma^0 = 0$. The agreement is excellent.
Note that the measured values for the initial and final momentum spread also
agree, while the skewness is underestimated by the numerical method (as
discussed in Section \ref{sec:model}).

\begin{figure*}[tb]
 \begin{center}
  \includegraphics[width=\textwidth, clip=true, trim=0 70 0 0]
  {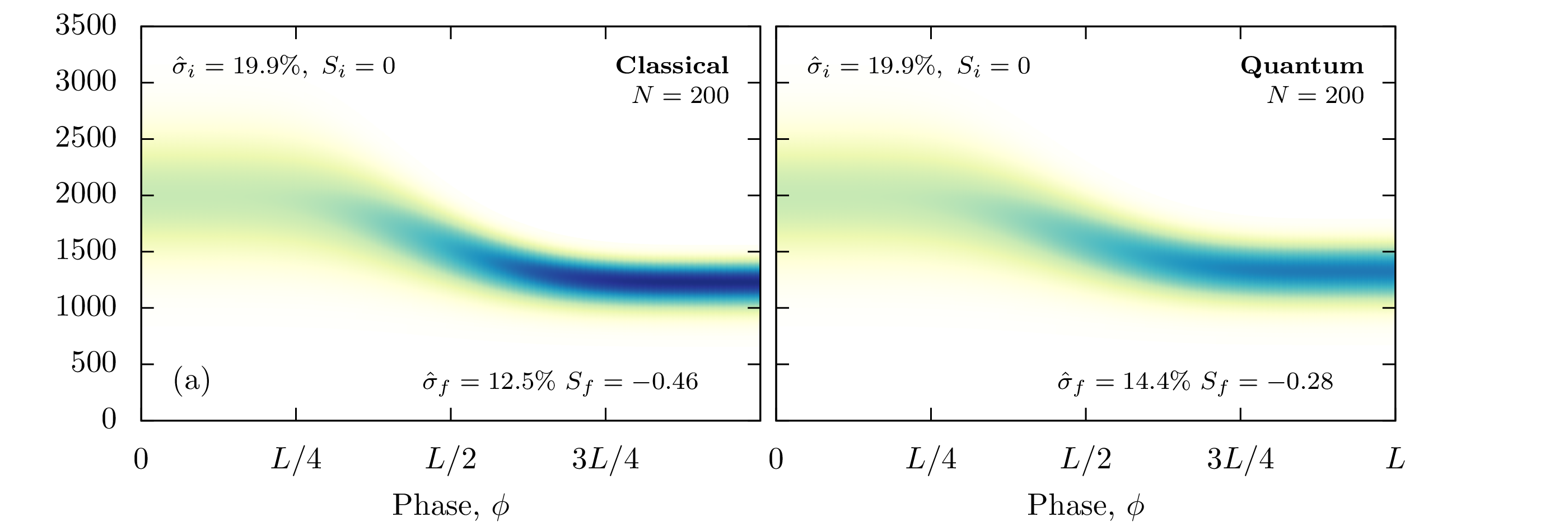}\\
  \vspace*{1em}
  \includegraphics[width=\textwidth, clip=true, trim=0 70 0 0]
  {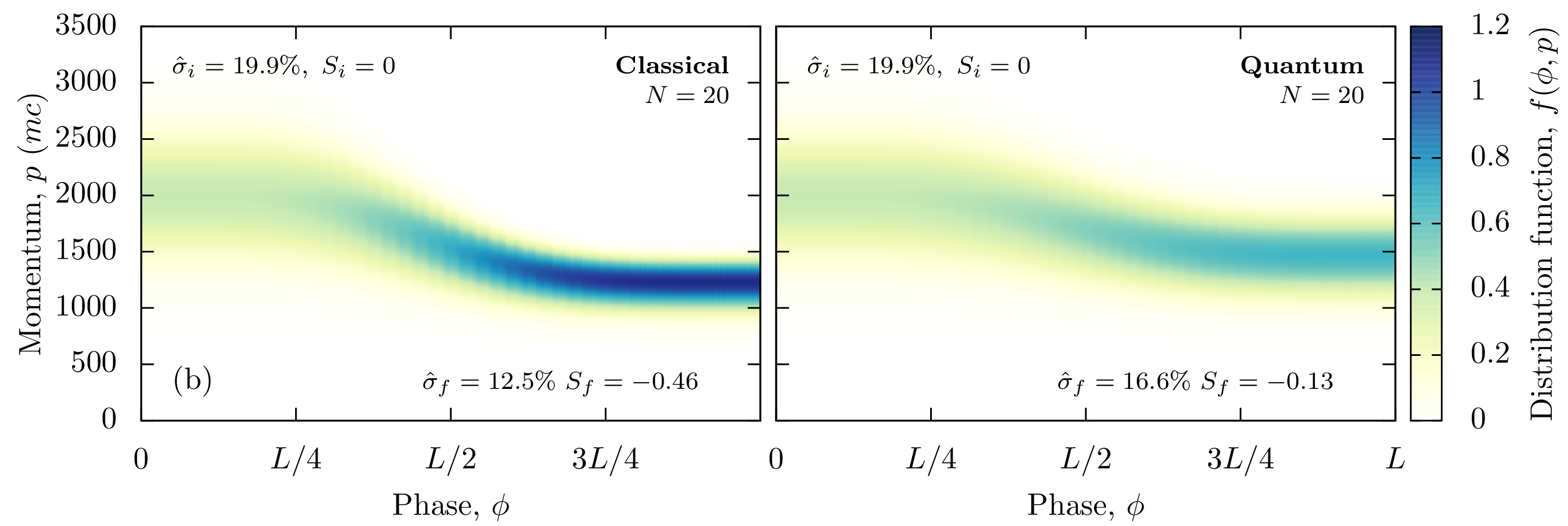}\\
  \vspace*{1em}
  \includegraphics[width=\textwidth]{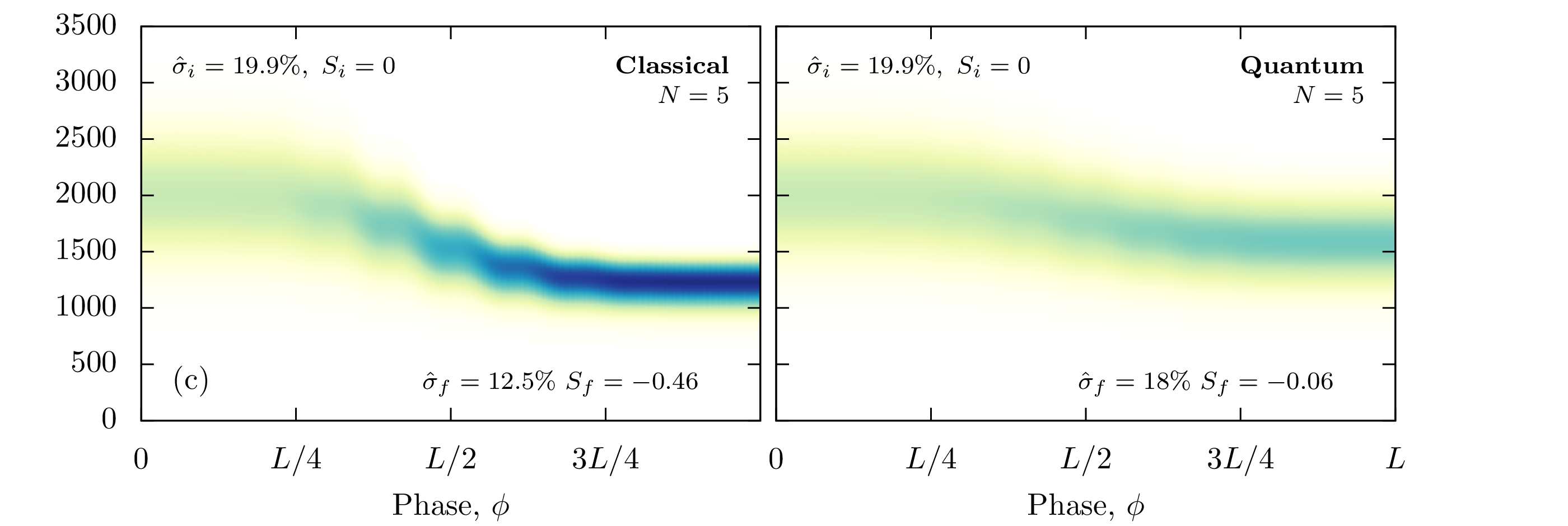}
 \end{center}
 \caption{(Colour online) The phase space evolution of the distribution
 function $f(\phi,p)$. Classical predictions are shown in the left-hand
 panels, while the corresponding semi-classical results are presented to the
 right. Values of the initial and final relative momentum spread and
 momentum skewness are displayed in each figure. The pulse length is reduced
 from $N=200$ cycles in part (a) to $N=20$ cycles in part (b), and finally
 to $N=5$ cycles in part (c). In each case, we observe an increase in the
 final mean momentum and its spread (reduction in beam cooling) predicted by
 the quantum model.}
 \label{fig:high_fluence}
\end{figure*}

Figure~\ref{fig:high_fluence} shows the variation of the particle
distribution on the $(\phi,p)$ phase space. As can clearly be seen in moving
from the classical Landau--Lifshitz theory (left) to the quantum model
(right), there are noticeable differences in the mean $\bar{\v}$, spread
$\hat{\sigma}$, and skewness $S$ of the distribution. We first note that the
deficit in measuring the initial $\hat{\sigma}_i = $ \SI\% $ < 20\%$ is due
to the finite number of particles used to represent the distribution (as
discussed at the end of Section \ref{sec:model} and illustrated in
Fig.~\ref{fig:recreate_dist}(b)).

For the classical theory, the final distribution only depends on the fluence
of the pulse, though this does not prevent the system from taking different
routes along the way. As the number of cycles is decreased, very different
intermediate behaviour is observed in Fig.~\ref{fig:high_fluence}, yet the
measured properties of the final distribution support this prediction: in
each case, we measure the mean momentum $\bar{\v}_f = \cp$ with a relative
spread $\hat{\sigma}_f = \cs\%$. This represents a significant contraction
of the phase space, where the average energy of the particle bunch decreases
significantly, as does its thermal spread (beam cooling), and the
distribution becomes more sharply peaked. In addition, we find the
development of a negatively-skewed distribution with $S_f = \cS$. In the
classical model this is readily understood, since the higher a particle's
momentum the more it radiates. This causes particles in the positive tail of
the distribution to be slowed down more than those in the negative tail.

The introduction of a semi-classical model in which the effect of radiation
reaction is reduced by the function $g(\chi)$ given by equation
\eqref{eq:QM_g} results in a reduction in the amount of phase space
contraction. Figure~\ref{fig:high_fluence}(a) for $N=200$ clearly
demonstrates this, with the final average momentum $\bar{\v}_f = \qap$ only
slightly higher than the classical case. The final relative momentum spread
is now \qas\%, showing that the final distribution is less sharply peaked.
While remaining negative, the skewness reduces in magnitude to $\qaS$,
because it is precisely the higher-energy particles (which were classically
most affected by radiation reaction) that now have this damping suppressed
due to larger $\chi$ (smaller $g(\chi)$).

These changes become more pronounced as we move to higher intensities (by
reducing the number of cycles). For $N=20$, as shown in
Fig.~\ref{fig:high_fluence}(b), we find that $\bar{\v}_f = \qbp$ and
$\hat{\sigma}_f = \qbs\%$ have both increased, with the skewness also
increasing to $S_f = \qbS$. This trend continues to $N = 5$ as shown in
Fig.~\ref{fig:high_fluence}(c). In this case, very little beam cooling
occurs for the quantum model, with the final relative momentum spread taking
the value $\hat{\sigma}_f = \qcs\%$ around $\bar{\v}_f = \qcp$. The profile
also remains much more Gaussian, with $S_f = \qcS$.

The reduction in phase space contraction (beam cooling) observed
here is in agreement with previous predictions \cite{Neitz2013}, which
provides further validation of our method for reconstructing the particle
distribution. Using this new method, it has been possible to investigate the
effects of the semi-classical model on a \emph{distribution} of particles.
The distributions are nicely reconstructed and do not feature any artefacts,
in contrast to other approaches \cite{Thomas2012}, which emphasises the
power of our method.

\begin{figure}[!t]
 \centering
 \includegraphics[width=0.49\textwidth]{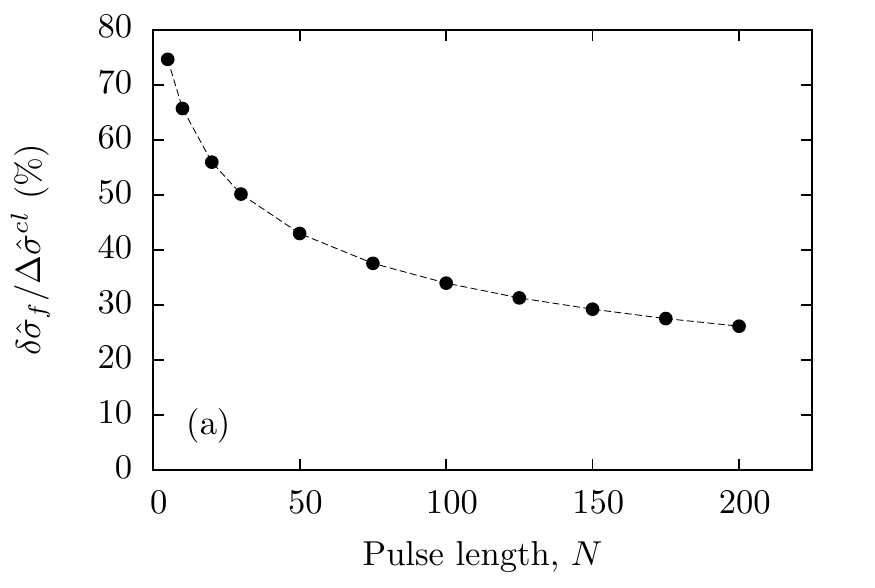}
 \includegraphics[width=0.49\textwidth]{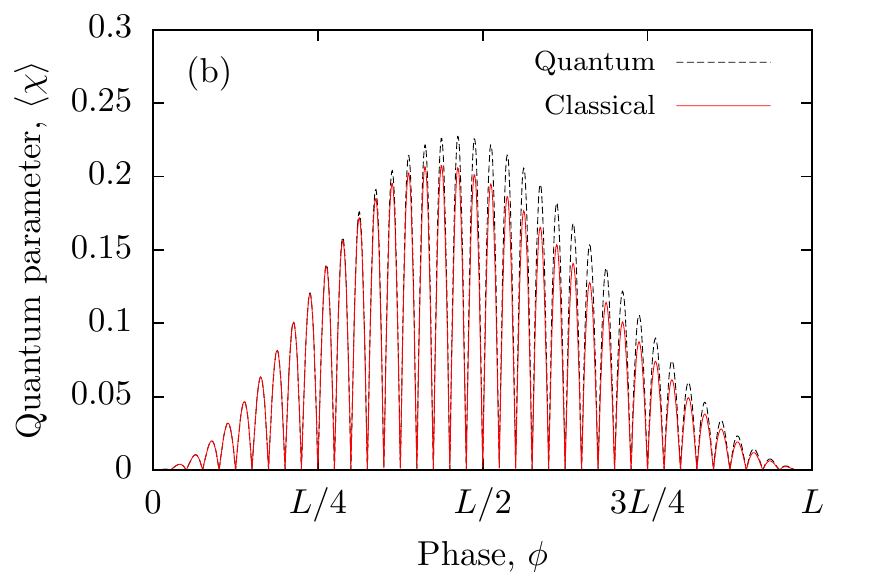}
 \caption{(Colour online) Part (a): variation of the final relative momentum
 spread difference $\delta\hat{\sigma}_f$ as a percentage of the total
 classical change, $\Delta\hat{\sigma}^\text{cl}$. The evolution of the
 average quantum parameter $\langle\chi\rangle$ according to the classical
 and semi-classical theories are plotted in part (b) for $N = 20$.}
 \label{fig:high_fluence_validity}
\end{figure}

Figure~\ref{fig:high_fluence} also shows how the difference between the
classical and quantum results increases as the intensity is increased (or as
$N$ is decreased). It is therefore interesting to consider the difference
$\delta\hat{\sigma}_f = \hat{\sigma}^\text{qm}_f - \hat{\sigma}^\text{cl}_f$
as a fraction of the total (constant) classical change in momentum spread,
$\Delta\hat{\sigma}^\text{cl} = \hat{\sigma}_i - \hat{\sigma}^\text{cl}_f$.
This can be found in Fig.~\ref{fig:high_fluence_validity}(a), where we see
that for $N=5$ the two predictions differ by about 75\%. As $N$ is
increased, this ratio is reduced because the average quantum parameter
$\langle\chi\rangle$ becomes smaller and radiation reaction is not so
heavily suppressed. It would be expected that the two models converge as
$N\to\infty$.

In cases where there is a large discrepancy between the predictions of the
two theories, it is especially important to be confident in the validity of
the model. As a semi-classical model, we expect it to remain valid into the
weakly quantum regime, such that particles experience \emph{instantaneous}
values $\chi^2 \ll 1$. Figure~\ref{fig:high_fluence_validity}(b) shows the
evolution of the bunch-average $\langle\chi\rangle$ as the bunch moves through a laser pulse
with $N=20$ according to the classical and semi-classical models. Initially,
there is good agreement between the two models, until the bunch approaches
the centre of the pulse, where the intensity becomes higher and the models
are significantly different. For completeness, we note that our highest
intensity case with $N=5$ satisfied $\langle\chi\rangle^2 < 0.22$.

\section{Conclusions}
\label{sec:conclusions}

The next few years will see the emergence of a number of new high-power
laser facilities operating at unprecedented field strengths, providing
access to fundamentally new physical regimes. This will allow us to
experimentally probe previously untested areas of physics, such as the
long-standing question of radiation reaction.

In this paper, we have analysed the transverse and longitudinal cooling of a relativistic electron beam as it interacts with an intense laser pulse, according to classical and semi-classical theories of radiation reaction. In the classical theory, we have found these two contributions to be equal, but quantum effects break this symmetry, leading to significantly less cooling in the longitudinal than the transverse directions.

To facilitate evaluation of the longitudinal beam cooling effects, we have introduced an innovative method to efficiently and
accurately calculate the distribution function for an electron beam
interacting with an intense laser pulse. This has been validated by comparison with an analytical solution to the Vlasov equation in the classical case, and used to compare
classical and quantum predictions of radiative cooling. We have found that quantum
effects can significantly alter the beam properties and, unlike the
classical case, can be influenced by the \emph{shape} of the laser pulse, not just its energy.

As we move into the quantum regime where final-state electron beam properties become
sensitive to pulse shape, it is becoming increasingly important to have an efficient
method in order to investigate the full parameter space. The approach developed here
to facilitate this study of beam dynamics provides a powerful tool
with wide-ranging application within the discipline.

The results presented in this paper are limited to the semi-classical case
$\chi^2 \ll 1$. However, it should be noted that, for the {\em longitudinal} beam cooling, this restriction is due
to the use of a deterministic equation of motion, and not the method
of sampling and reconstructing the distribution. There should be no
obstruction to exploring more strongly quantum regimes (such as higher initial
beam energies $\sim5$ GeV available at ELI) using this approach with a
stochastic equation where photon emission probabilities are determined by
strong field QED, as in \cite{Elkina2011,Green2014}. This will be addressed in future
work, along with an investigation of stochastic {\em transverse} beam cooling.

\section*{Acknowledgements}
This work is supported by the UK EPSRC (Grant EP/J018171/1); the ELI-NP
Project; and the European Commission FP7 projects Laserlab-Europe (Grant
284464) and EuCARD-2 (Grant 312453). Dataset available online (DOI:
10.15129/79f9c58d-7a43-4cc0-a613-ebc028519e5b).

\section*{References}
\bibliographystyle{iopart-num}
\bibliography{refs}

\end{document}